\documentclass[aps,pra,twocolumn,epsf,showpacs,superscriptaddress]{revtex4}
\usepackage{graphicx}
\usepackage{amsmath,amssymb,latexsym}
\usepackage{bm}

\newcommand{\bra}[1]{\langle #1|}
\newcommand{\ket}[1]{|#1\rangle}
\newcommand{\braket}[2]{\langle #1|#2\rangle}
\newcommand{\pton}[1]{\left(#1\right)}
\newcommand{\pqua}[1]{\left[#1\right]}

\newcommand{\der}[2]{\frac{\partial #1}{\partial #2}}

\begin{document}

\title{Mirror-induced decoherence in hybrid quantum-classical theory} 

\author{Aniello Lampo} 
\affiliation{ICFO, Mediterranean Technology Park, Av. C.F. Gauss 3, 
08860 Castelldefels (Barcelona), Spain}
\email{aniello.lampo@icfo.es}
\author{Lorenzo Fratino} 
\affiliation{SEPnet \& Hubbard Theory Consortium, Department of Physics, Royal Holloway, University of London, Egham, Surrey TW20 0EX, UK}
\email{lorenzo.fratino.2013@live.rhul.ac.uk}
\author{Hans-Thomas Elze}
\affiliation{Dipartimento di Fisica ``Enrico Fermi'',  Largo Pontecorvo 3, I-56127 Pisa, Italia} 
\email{elze@df.unipi.it}

\begin{abstract} 
We re-analyse the optomechanical interferometer experiment proposed by 
{\it Marshall, Simon, Penrose and Bouwmeester} with the help of a recently developed 
quantum-classical hybrid theory. This leads to an alternative evaluation 
of the mirror induced decoherence. Surprisingly, we find that it behaves 
essentially in the same way for suitable initial conditions and experimentally 
relevant parameters, no matter whether the mirror is considered a classical or 
quantum mechanical object. We discuss the parameter ranges where this result 
holds and possible implications for a test of spontaneous collapse models, for which 
this experiment has been designed.  
\end{abstract}
\maketitle 

\section{Introduction} 
The optomechanical experiment proposed by {\it Marshall et al.} \cite{Penrose} 
presents one of the first systems where one attempts to produce and detect 
coherent quantum superpositions of spatially separated macroscopic states, and to test 
various decoherence and wave function collapse models 
\cite{Diosi84,Diosi1,Penrose98,Bassi1,Diosi,decoherence}. 

The basic idea here is close in spirit to Schr\"{o}dinger's original discussion 
\cite{Bose1,Bose2}: a microscopic quantum system (photon), for which the 
superposition principle is undoubtedly valid, is coupled with a macroscopic object (mirror), 
in order to transfer interference effects from the former to the latter, creating a macroscopic superposition state. For this goal, one employs a Michelson interferometer with a tiny 
moveable mirror in one arm. In this way, since the photon displaces through its radiation pressure the tiny mirror, the initial superposition of the photon being in either arms 
causes the system to evolve into a superposition of states corresponding to two distinct locations of the mirror. 

Nevertheless, before being able to detect macroscopic superpositions, a serious obstacle 
has to be overcome: decoherence induced by the mirror itself on the photon. The photon, indeed, 
cannot be dealt with as an isolated system, because it interacts with the mirror and, hence, 
decoherence can occur destroying any photon coherent superposition \cite{Zurek,MSR}.
In this case, no interference effects can be transferred to the mirror. 

Mirror induced decoherence has been examined in Ref.\,\cite{Penrose}, considering both, the 
mirror and the photon, as quantum objects. However, the size of the former ($\approx1\mu$m) far exceeds the scales which are typical of the explored quantum regime. Therefore, 
a classical description of the mirror should be investigated as an alternative 
and differences or similarities with a quantum one need to be confronted with the planned 
experiments. 

The purpose of our paper is to study the decoherence process with the mirror treated 
as a classical, rather than a quantum subsystem, while the photon obviously retains its 
quantum nature. Thus, we have to deal with a model comprising a quantum and a classical sector, 
which coexist and interact. Such a situation needs a particular theoretical framework  
for a consistent description, namely a quantum-classical hybrid theory. 

There has been much interest in hybrid theories, both for practical and theoretical reasons. 
From a theoretical point of view, hybrid theories have originally been devised to provide a different approach to the quantum measurement problem \cite{Sherry}.
Furthermore, a quantum-classical hybrid theory may be employed to describe consistently the interaction between quantum matter and classical spacetime \cite{Bou}.  
See also, for example, the related studies in 
Refs.\,\cite{CaroSalcedo99,DiosiGisinStrunz,PeresTerno,HallReginatto05,ZhangWu06,Hall08,Manko12}. 

Even if one is not inclined to modify certain ingredients of quantum theory, there is also 
clearly practical interest in various forms of hybrid dynamics, in particular in nuclear, 
atomic, or molecular physics.
The Born-Oppenheimer approximation, for example, is based on a separation of interacting slow 
and fast degrees of freedom of a compound object.
The former are treated as approximately classical, the latter as of quantum mechanical nature.
Moreover, mean field theory, based on the expansion of quantum mechanical variables into a classical part plus quantum fluctuations, leads to another approximation scheme and another 
form of hybrid dynamics.
This has been reviewed more generally for macroscopic quantum phenomena in Ref.\,\cite{Gnac}.
In all these cases hybrid dynamics is considered as an approximate description of an intrinsically quantum mechanical object.
Such considerations are and will become increasingly important for the precise manipulation of quantum mechanical objects by apparently and for all practical purposes classical means, especially in the mesoscopic regime. 

In particular, we recall the hybrid theory elaborated in \cite{Elze1,Elze2,Elze3,Elze4}, 
which overcomes the known impediments found in earlier work. For a closely related approach, 
see also Ref.\,\cite{Buric}.
Herein, the classical sector is described by the standard analytical classical mechanics, 
while the description of the quantum sector is based on Heslot's representation \cite{Heslot}, 
{\it cf.} Ref.\,\cite{Strocchi}, allowing to express quantum mechanics in a Hamiltonian 
framework. 
Similarly as in classical physics, it is then possible to present states of the quantum sector 
in terms of couples of real time-dependent functions, rather then vectors, which play 
the role of canonical variables. Furthermore, the observables are no longer given by 
self-adjoint operators, but are represented by real quadratic functions of these canonical 
variables. In this way, an entire hybrid system can be studied in one, uniform, scheme. 

In Section\,$\ref{HH}$, we will employ this hybrid theory to specify a Hamiltonian, which 
constitutes the starting point for the analysis of the dynamics of the 
{\it Marshall et al.} optomechanical system. -- In Section\,$\ref{EM}$, we will derive 
and solve the corresponding equations of motions. -- The solutions of these equations will 
be employed to calculate the off-diagonal matrix elements of the reduced density matrix 
for the photon and allow the evaluation of the decoherence induced by the classical mirror.  
The analysis of this decoherence process and the discussion of the results obtained 
will be presented in Section\,$\ref{MIDec}$. -- 
Finally, in Section\,$\ref{PIP}$, we will relate the off-diagonal matrix elements to 
quantities that can be determined experimentally. -- In the concluding section, 
implications of our results shall be discussed.  

\section{The quantum-classical hybrid Hamiltonian}\label{HH}
The Hamiltonian for the optomechanical interferometer system proposed by 
{\it Marshall et al.} \cite{Penrose} consists naturally of three different parts  
which have their correspondents in our hybrid description: 
terms related to the photon (the quantum sector), terms associated with the mirror 
(the classical sector), and the hybrid coupling of both sectors.  

Since the photon is treated quantum mechanically, it is represented by a Hamilton  
operator $\hat{H}_{QM}$:
\begin{equation}\label{HQM}
\hat{H}_{QM}=\hbar\omega (\hat{c}^{\dagger}_{A}\hat{c}_{A}
+\hat{c}^{\dagger}_{B}\hat{c}_{B}) 
\;\;, \end{equation}
where $\hat{c}^{\dagger}_{A}$ and $\hat{c}_{A}$, respectively, are creation and 
annihilation operators for a photon in arm A, and correspondingly for arm B. -- 
Instead, the mirror is considered here as a classical subsystem. While it was described  
as a quantum harmonic oscillator in Ref.\,\cite{Penrose}, it is now represented 
by a classical one with Hamiltonian $H_{CL}$:
\begin{equation}\label{HCL}
H_{CL}=\frac{p^{2}}{2M}+\frac{M\Omega^{2}}{2}x^{2} 
\;\;, \end{equation}
in which $x$ and $p$ denote position and momentum of the mirror, respectively. -- 
The hybrid coupling $\hat{I}$, {\it i.e.} the interaction between photon and mirror, 
incorporates both, a quantum operator and a classical variable, for photon and mirror, 
respectively. Being essentially related to the radiation pressure of the photon, as shown  
in detail in Refs.\,\cite{Man1,Man2,Pace,Law1,Law2}, we have:
\begin{equation}\label{HIT}
\hat{I}=\hbar gx\hat{c}^{\dagger}_{A}\hat{c}_{A} 
\;\;,\end{equation}
where $g:=\omega /L$.

Following Refs.\,\cite{Elze1,Elze2,Elze3,Elze4}, we obtain the full hybrid Hamiltonian $H$ as:
\begin{equation}\label{FHH}
H=H_{CL}+\bra{\psi}(\hat{H}_{QM}+\hat I)\ket{\psi} 
\;\;, \end{equation}
in which $\ket{\psi}$ denotes a generic photon state, satisfying always $\braket{\psi\pton{t}}{\psi\pton{t}}=1$\,. 
 
In order to evaluate the expectation values in Eq.\,(\ref{FHH}),  
we consider the orthonormal basis in the Hilbert space of the photon formed by the  
two vectors $\ket{1,0}$ and $\ket{0,1}$, representing the states in which the photon 
is either in arm A or B of the interferometer; since we consider only 
one-photon states, this is also complete. Accordingly, we expand the state $\ket{\psi}$ 
as follows:
\begin{equation}\label{Expansion}
\ket{\psi}=\frac{1}{\sqrt{2\hbar}}\pton{X_{A}+iP_{A}}\ket{1,0}
+\frac{1}{\sqrt{2\hbar}}\pton{X_{B}+iP_{B}}\ket{0,1} 
\;\;, \end{equation}
where $X_{A}$, $X_{B}$, $P_{A}$, and $P_{B}$ are real time-dependent functions, 
which play the role of canonical variables \cite{Elze1,Elze2,Elze3,Elze4}. This gives: 
\begin{eqnarray} 
H&=&\frac{p^2}{2M}+\frac{M\Omega^2}{2}x^2
\nonumber \\ [1ex] \label{FinHH} 
&\;&+\frac{\omega}{2}\sum_{i=A,B}\pton{X^2_{i}+P^2_{i}}  +\frac{g}{2}x\pton{X^2_{A}+P^2_{A}}
\;\;. \end{eqnarray}
This is the full hybrid Hamiltonian for the {\it Marshall et al.} optomechanical system, 
which is a real function of position and momentum of the mirror and of the canonical variables 
pertaining to the photon, resembling a completely classical formalism.
Nevertheless, the quantum nature of the photon is described correctly according to the 
representation of Eq.\,(\ref{Expansion}). 

\subsection{Comments} 
The scenario  \cite{Penrose} on which our work is based, resulting 
in the hybrid Hamiltonian (\ref{FinHH}), may be related 
to real experiments.  We comment on two important aspects.   

Assuming that the {\it single photon within} the interferometer typically will be produced 
by a source {\it outside} of the cavity under consideration, the interferometer 
should be treated as an open system. This has been pointed out in Ref.\,\cite{Chen1}  
and studied in a fully quantum mechanical approach (invoking the rotating wave 
approximation when coupling photon cavity mode inside and continuum modes 
outside).  Depending on the bandwidth of the initial photon wave function as 
compared to the cavity linewidth, the photon  may obtain simultaneously nonzero amplitudes 
to be inside and outside of the cavity. -- This will not affect the hybrid coupling 
(\ref{HIT}), on which we have presently focussed attention. However, 
it introduces an additional time dependence into the dynamics. Our description of the photon sector can naturally be adapted to this, as before \cite{Chen1}, while the  
coupling to the considered classical mirror remains the same.  Anticipating the results 
obtained in the following, we expect similar effects as observed in Ref.\,\cite{Chen1}, 
which remains to be confirmed by a detailed study. 

Furthermore, one wonders whether our present derivation must be generalized, in 
order to allow for the simultaneous presence of {\it several photons} inside the cavity, 
which might be experimentally of interest. Let us consider for illustration the 
interferometer as a closed system and assume a two-photon state inside. Each photon 
can be present in its arm A or B. Any resulting two-photon state can be embedded in a 
four-dimensional Hilbert space (e.g. spanned by Bell states) and be 
expanded analogously to Eq.\,(\ref{Expansion}) above. (This has recently been applied 
in a somewhat similar setting of two q-bits interacting with a classical oscillator 
\cite{Lorenzo}.) Correspondingly, there will be a proliferation of terms in the resulting 
hybrid Hamiltonian. Nevertheless, the bilinearity of the hybrid Hamiltonian 
in the canonical coordinates, as in Eq.\,(\ref{FinHH}),  will not be affected. 
The character of the eventually resulting coupled equations of motion, 
cf. Section\,$\ref{EM}$, will only change by  number. 
However, for high intensity of the cavity field, genuinely nonlinear multi-photon 
interactions are to be expected, rendering the hybrid coupling constant $g$ in 
Eq.\,(\ref{HIT}) effectively  time and photon state dependent. This is clearly an 
interesting regime to be studied, concerning comparison between 
fully quantum mechanical and hybrid descriptions in particular.  

This leads us to mention another hybrid system currently attracting much interest 
\cite{Carlip,BassiEtAl}, 
namely a massive body (considered quantum mechanical) interacting with the 
gravitational field (considered classical and nonrelativistic), thus 
interchanging the role of quantum mechanical and classical degrees freedom as 
compared to the situation presently studied. It has recently been shown that 
this hybrid can be represented by the nonlinear Schr\"odinger-Newton equation, in the 
limit that the extended body is composed of many ``mass-concentrating'' constituents 
(atoms or similar)  \cite{Chen2}. 

It remains to be further examined to what extent such effectively {\it nonlinear} 
descriptions of quantum-classical hybrids are consistent in the sense of criteria 
discussed earlier \cite{CaroSalcedo99,DiosiGisinStrunz,PeresTerno,Hall08,Elze1,Elze2,Elze3,Elze4,Buric}.  

\section{The hybrid equations of motion}\label{EM} 
Having specified the hybrid Hamiltonian for the {\it Marshall et al.} optomechanical 
interferometer system, we are now ready to derive the corresponding equations of motion. 
Following the procedure defined by the appropriate Poisson bracket structure for 
the algebra of observables and the Hamiltonian, in particular, we will obtain the 
equations of motion and their solutions here, {\it cf.} Refs.\,\cite{Elze1, Elze2, Elze3,Elze4}.

\subsection{Derivation of the equations}
We begin with the photon in arm A. Its equations of motion are formally given by:
\begin{equation}\label{AEQ1}
\der{X_{A}}{t}=\{X_{A},H\}_{X}\quad,\quad\der{P_{A}}{t}=\{P_{A},H\}_{X}
\;\;, \end{equation}
with $\{f,g\}_{X}:=\{f,g\}_{CL}+\{f,g\}_{QM}$. Here the ``quantum mechanical'' 
Poisson bracket is given by 
$\{f,g\}_{QM}:=\sum_{i=A,B}\pton{\der{f}{X_{i}}\der{g}{P_{i}}-\der{g}{X_{i}}\der{f}{P_{i}}}$,  
while the classical Poisson bracket $\{f,g\}_{CL}$ is the usual one, of course -  
following the rules of the hybrid theory constructed in Ref.\,\cite{Elze1}. 
Inserting Eq.\,(\ref{FinHH}) into Eq.\,(\ref{AEQ1}), we obtain:
\begin{equation}\label{AEQ2}
\der{X_{A}}{t}=\omega P_{A}-gxP_{A}\;\; ,\;\;\; 
\der{P_{A}}{t}=-\omega X_{A}+gxX_{A}
\;\;. \end{equation}
In the same manner, we obtain the equations of motion for the photon in arm B:
\begin{equation}\label{BEQ}
\der{X_{B}}{t}=\omega P_{B}\;\; ,\;\;\; 
\der{P_{B}}{t}=-\omega X_{B} 
\;\;. \end{equation}
Finally, for the mirror, we similarly obtain:
\begin{equation}\label{MEQ}
\der{x}{t}=\frac{p}{M}\;\; ,\;\;\; 
\der{p}{t}=-M\Omega^2x+\frac{g}{2}\pton{X_{A}^2+P_{A}^2} 
\;\;. \end{equation}
Remarkably, the equations of motion for our quantum-classical hybrid system have 
a completely classical appearance. This will be employed in the following 
derivation of their solutions. 

\subsection{Solution of the hybrid equations}
The equations obtained in the previous subsection describe a set of coupled 
harmonic oscillators, which can be solved analytically. 
In particular, while the equations for the photon in arm B are completely 
decoupled, those associated with the mirror and the photon in arm A are coupled.
This is as expected, since the photon in arm B does not interact with anything, 
while when it is in arm A it inevitably interacts with the mirror. 

The coupling term in Eq.\,(\ref{MEQ}), $\propto\pton{X_{A}^2+P_{A}^2}$, at first sight, 
complicates solving of the equations. However, 
this term consists in a constant of motion and, therefore, can be replaced by its 
initial value. We note that: 
\begin{equation}\label{Const}
\frac{1}{2}\der{\pton{X^2_{A}+P^2_{A}}}{t}=X_{A}\der{X_{A}}{t}+P_{A}\der{P_{A}}{t}=0 
\;\;, \end{equation}
using Eq.\,(\ref{AEQ2}). In order to determine the value of this constant of motion, 
we consider its physical meaning: it simply describes the probability to find the photon in 
arm A, expressed in the representation of Eq.\,(\ref{Expansion}). 
We assume a fifty-fifty beam splitter in the interferometer \cite{Penrose}. Therefore, 
this probability is equal to $1/2$:
\begin{eqnarray}
\frac{1}{2}&=&\bra{\psi}\hat{c}^{\dagger}_{A}\hat{c}_{A}\ket{\psi}
=\frac{1}{2\hbar}\pton{X^2_{A}+P^2_{A}}\bra{0,1}\hat{c}^{\dagger}_{A}\hat{c}_{A}\ket{1,0}
\nonumber \\ [1ex] \label{norma} 
&=&\frac{1}{2\hbar}\pton{X^2_{A}+P^2_{A}}\braket{0,1}{1,0}=
\frac{1}{2\hbar}\pton{X^2_{A}+P^2_{A}}
, \end{eqnarray}
since $\langle 0,1|=|1,0\rangle^\dagger$. Thus, we have:
$X^2_{A}+P^2_{A}=\hbar$\,.      
Similarly, we find: 
$X^2_{B}+P^2_{B}=\hbar$\,.  

Making use of this in Eq.\,(\ref{MEQ}), the equations for the mirror become:
\begin{equation}\label{SM2}
\der{x}{t}=\frac{p}{M}\;\; ,\;\;\; \der{p}{t}=-M\Omega^2x+\frac{\hbar g}{2}
\;\;, \end{equation}
which can be easily solved to give:
\begin{eqnarray}\label{xpmirr}
x\pton{t}&=&A\sin\pton{\Omega t+\phi}-\frac{\hbar g}{2M\Omega^2}\;\; , 
\\ [1ex] \label{pxmirr} 
p\pton{t}&=&AM\Omega \cos\pton{\Omega t+\phi}
\;\;, \end{eqnarray}
where $A$ and $\phi$, respectively, represent amplitude and  
phase of the oscillation of the mirror, which are determined by the initial conditions. -- 
With:
\begin{equation}\label{xp0}
x\pton{0}=A\sin\pton{\phi}-\frac{\hbar g}{2M\Omega^2}
\;\; ,\;\;\; 
p\pton{0}=AM\Omega \cos\pton{\phi}   
\;\;, \end{equation}
and abbreviating $x_0\equiv x\pton{0}$, $p_0\equiv p\pton{0}$, we find: 
\begin{eqnarray}\label{Am}
A\pton{x_{0},p_{0}}&=&
\frac{p_{0}}{M\Omega\cos \big [\phi (x_0,p_0)\big ]} 
\;\;, \\ [1ex] \label{Phim} 
\phi\pton{x_{0},p_{0}}&=&
\arctan \big [(M\Omega x_{0}+\frac{\hbar g}{2\Omega})/p_0\big ] 
\;\;. \end{eqnarray} 

Knowing the solutions of the equations for the mirror, we can solve those for the photon in 
arm A, Eqs.\,(\ref{AEQ2}). It is straightforward to obtain:
\begin{eqnarray}\label{XAsol} 
\frac{X_{A}\pton{t}}{\sqrt\hbar}=\cos\big (\Omega_+t
+\frac{A g}{\Omega}\big [\cos\pton{\Omega t+\phi}-\cos\pton{\phi}\big ]\big )
,\; \\ [1ex] \label{PAsol}
\frac{P_{A}\pton{t}}{\sqrt{\hbar}}=-\sin\big (\Omega_+ t
+\frac{Ag}{\Omega}\big [\cos\pton{\Omega t+\phi}-\cos\pton{\phi}\big ]\big )
,\; \end{eqnarray}
conveniently introducing $\Omega_+$ and $k$,   
$\Omega_+:=\omega+k^{2}\Omega :=\omega +\hbar g^2/(2M\Omega^2)$\,. 

Finally, regarding the photon in arm B, we find: 
\begin{equation}\label{FinalB}
X_{B}\pton{t}=\sqrt{\hbar}\cos\pton{\omega t}\quad,\quad P_{B}\pton{t}=-\sqrt{\hbar}\sin\pton{\omega t}
\;\;. \end{equation}
Here we recalled the normalization of the photon state for arm B, which follows similarly 
as in Eq.\,(\ref{norma}), in order to fix the amplitudes. 
The phase, instead, is determined by choosing:  
$X_{B}\pton{0}=1$ and $P_{B}\pton{0}=0$\,.

\section{The mirror induced decoherence}\label{MIDec}
In this section, we report quantitative results concerning the decoherence induced by 
the classical mirror on the quantum photon. 

The relevant information is contained in the off-diagonal elements of the 
reduced density matrix for the photon. Presently, these matrix elements are given by:
\begin{equation}\label{odme1}
\rho_{AB}=\braket{1,0}{\psi}\braket{\psi}{1,0}=\rho_{BA}^{\; *} 
\;\;, \end{equation} 
with $\langle 1,0|=|0,1\rangle^\dagger$. 
Inserting Eq.\,(\ref{Expansion}) here, we obtain:
\begin{equation}\label{odme2}
\rho_{AB}\pton{t}=\frac{1}{2\hbar}\pton{X_{A}+iP_{A}}\pton{X_{B}-iP_{B}}
\;\;, \end{equation}
and recalling Eqs.\,(\ref{XAsol}), (\ref{PAsol}), and (\ref{FinalB}), this becomes:
\begin{eqnarray}
&\;&\rho_{AB}\pton{t;x_{0},p_{0}}=\frac{1}{2}\exp\big\{i\omega t\big\} 
\nonumber \\ [1ex] \label{finalrho}
&\cdot&\exp\big\{-i\big [\Omega_+ t
+\frac{Ag}{\omega}[\cos\pton{\Omega t
+\phi}-\cos\pton{\phi}]\big ]\big\} 
,\;\; \end{eqnarray}
where the dependence on $x_{0}$ and $p_{0}$ is through $A$ and $\phi$. 

Taking the modulus of the matrix element $\rho_{AB}$, we 
obtain the mirror induced decoherence as a function of time, {\it i.e.} the visibility 
of interference of the photon. Since the result of Eq.\,(\ref{finalrho}) is a pure phase, 
its modulus is a constant:
\begin{equation}
|\rho_{AB}|=\frac{1}{2} 
\;\;. \end{equation}
Thus, for pointlike initial conditions in the classical phase space of the mirror, 
no mirror induced decoherence occurs. In this case, the photon remains in its initial pure 
state, the coherent superposition of being in either of the arms of the Michelson 
interferometer, and the corresponding interference effects are preserved. 

We emphasize that this result concerns the situation where the initial conditions of the 
classical mirror are perfectly known and are represented by a point in phase space. 
It shows a clear and expected difference between the present hybrid and the purely quantum 
approach \cite{Penrose}. The latter results in   
mirror induced decoherence, even before thermal averaging. 

However, suppose that there is some loss of information consisting in somewhat imprecisely 
known initial position and momentum of the mirror. Instead of sharp initial conditions, 
we may have a probability distribution over phase space. -- 
For instance, we consider a Boltzmann distribution:
\begin{equation}\label{MB}
f\pton{x_{0},p_{0}}:=\frac{\beta\Omega}{2\pi}
\exp\big\{-\beta\pton{\frac{p^{\; 2}_{0}}{2M}+\frac{M\Omega^2x^{\; 2}_{0}}{2}}\big\}
\;\;, \end{equation}
depending on the inverse temperature, $\beta :=1/k_BT$.  
In this case, the physically relevant matrix element is given by the thermal average 
of the result of Eq.\,(\ref{finalrho}):
\begin{equation}\label{rhof}
<\rho_{AB}>_{f}=\int\rho_{AB}\pton{t;x_{0},p_{0}}f\pton{x_{0},p_{0}}\mbox{d}x_{0}\mbox{d}p_{0} 
\;\;. \end{equation} 

In order to calculate this integral, we rewrite it more conveniently, 
using the abbreviations 
$\kappa^2:=\hbar g^2/(2M\Omega^3)$ (previously introduced in Ref.\,\cite{Penrose}), 
$\;\theta\pton{t}:=\Omega t-\sin\pton{\Omega t}$\,, and:   
\begin{eqnarray}\label{intg}
h_1\pton{x_{0},t}&:=&\frac{\beta M\Omega^2}{2}x^{\; 2}_{0}
+i\frac{g}{\Omega}\sin\pton{\Omega t}x_{0} 
\;\;, \\ [1ex] \label{inth}
h_2\pton{p_{0},t}&:=&-\frac{\beta}{2M}p^{\; 2}_{0}
-i\frac{g}{M\Omega^2}\big [\cos\pton{\omega_{m}t}-1\big ]p_{0}
.\;\;\; \end{eqnarray}
Thus, we obtain from Eq.\,(\ref{rhof}): 
\begin{eqnarray} 
&\phantom .&<\rho_{AB}\pton{t}>_{f}\;=\;
\frac{\beta\Omega}{4\pi}\exp\{ -i\kappa^2\theta\pton{t} \} 
\nonumber \\ [1ex] \label{rhof1}
&\phantom .&\;\;\cdot 
\int^{\infty}_{-\infty} \exp\{ h_1\pton{x_{0},t}\} \mbox{d}x_{0} 
\int^{\infty}_{-\infty} \exp\{ h_2\pton{p_{0},t}\} \mbox{d}p_{0}
\nonumber \\ [1ex] \label{rhof2}
&=&\frac{1}{2}\exp\big\{ -i\kappa^2 [\Omega t-\sin\pton{\Omega t}] 
-z_{CL}^2[1-\cos\pton{\Omega t}]\big\}  
,\;\;\;\;\;\; \end{eqnarray} 
with $z_{CL}^2:=g^2/(\beta M\Omega^4)=2\kappa^{2}k_BT/(\hbar\Omega )$.

In order to evaluate the mirror induced decoherence, we calculate the modulus of 
this averaged matrix element (commonly referred to as {\it ``visibility''}):
\begin{equation}\label{mod}
|<\rho_{AB}\pton{t}>_{f}|=\frac{1}{2}\exp\big\{-z_{CL}^2[1-\cos\pton{\Omega t}]\big\}  
\;\;. \end{equation}
This shows the decoherence induced by the classical mirror on the quantum photon, 
taking into account that the mirror initial conditions are thermally distributed. 

\begin{figure}[!htpb]
\begin{center}
\includegraphics[width=7.3cm, height=7.5cm, keepaspectratio]{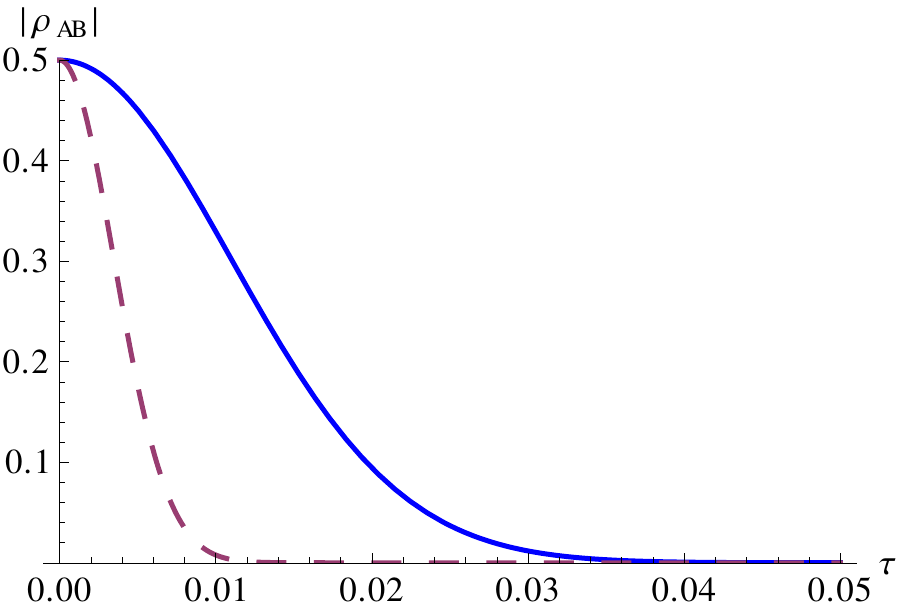}
\includegraphics[width=7.3cm, height=7.5cm, keepaspectratio]{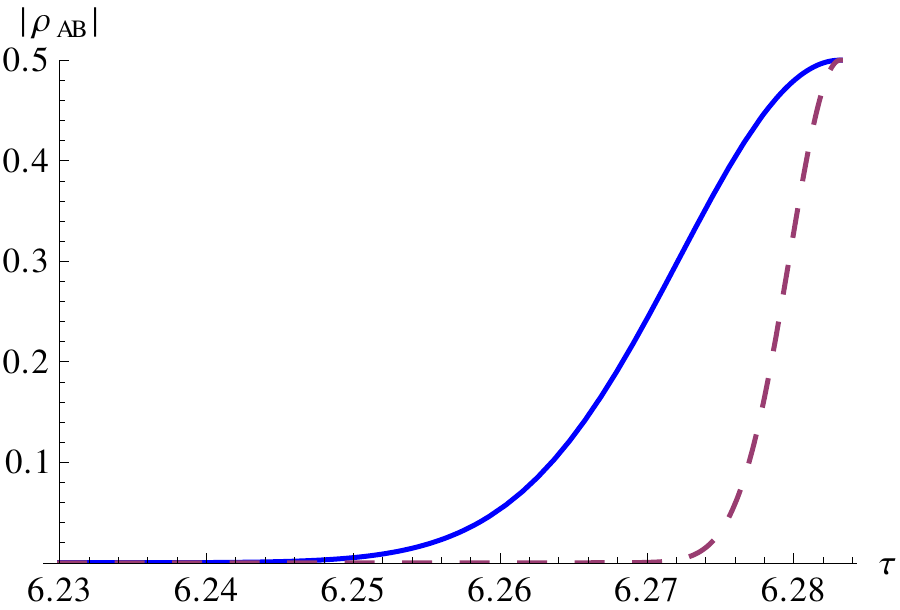}
\caption{\label{mid} $|<\rho_{AB}\pton{t}>_{f}|$ as a function of $\tau :=\Omega t$, for $\kappa =1$, 
$\Omega=2\pi\cdot 500$Hz. 
The dashed line (purple) is for $T=10^{-3}$K, the full line (blue) for $T=10^{-4}$K.}
\end{center}
\end{figure}

The Fig.\,\ref{mid} shows the temporal behaviour of the visibility of interference. 
As in the purely quantum case, the visibility has a maximum at the initial instant 
and then decreases, due to the mirror induced decoherence.
However, after half a period of the mirror oscillation, we observe a revival of 
coherence of the photon and the visibility returns to its maximum value exactly at 
$t=2\pi /\Omega$. 

It is important to realize that the result of Eq.\,(\ref{mod}), incorporating the 
thermal average over classical initial conditions, has {\it exactly}  
the same form as the earlier one obtained in Ref.\,\cite{Penrose} 
for a quantum mechanical mirror as part of the interferometer. 
In particular, we find the {\it same time dependence}. 
The only difference resides in that our parameter $z_{CL}$, defined after 
Eq.\,(\ref{rhof2}), 
has to be replaced by the corresponding parameter $z_{QM}$ given by: 
\begin{eqnarray}  
z_{QM}^2&=&2\kappa^2\Big (\bar n(\hbar\Omega /k_BT)+1/2\Big )
\nonumber \\ [1ex] \label{zrel}
&=& 
z_{CL}^2(\hbar\Omega /k_BT)\Big (\bar n(\hbar\Omega /k_BT)+1/2\Big )  
\;\;, \end{eqnarray} 
with the Bose-Einstein distribution $\bar n(x):=(\exp (x)-1)^{-1}$. 
Here we incorporated the appropriate finite temperature correction indicated 
(but not explicitly  given) in Ref.\,\cite{Penrose} for the quantum mechanical mirror. 

 Thus, we find that in the {\it high-temperature limit}, with $\hbar\Omega /k_BT\ll 1$, 
both parameters coincide, 
\begin{equation}\label{hT} 
z_{QM}^2=z_{CL}^2\Big (1+\frac{1}{12}[\hbar\Omega /k_BT]^2
+\mbox{O}([\hbar\Omega /k_BT]^4)\Big ) 
\;\;, \end{equation} 
and, consequently,  the visibilities given by the right-hand side of Eq.\,(\ref{mod}), with 
either $z_{CL}$ or $z_{QM}$ inserted, become equal as well, for all times! 

More generally, 
considering the ratio $\eta$ of the result of Eq.\,(\ref{mod}) divided by the 
quantum mechanical result from \cite{Penrose}, 
$\eta :=|<\rho_{AB}\pton{t}>_{f}|/|<\rho_{AB}\pton{t}>_{QM}|$, 
we find numerically that 
-- for experimentally relevant temperatures 
$10^{-6}\mbox{K}<T<10^{-3}\mbox{K}$ and mirror frequency 
$\Omega=2\pi\cdot 500$Hz -- the 
deviation of both results can be correspondingly bounded by 
$10^{-2}>\eta -1>10^ {-6}$, indeed a surprising result. Furthermore, 
due to identical time dependence, $\propto 1-\cos (\Omega t)$, in the 
exponent on the right-hand side of Eq.\,(\ref{mod}) and the corresponding 
quantum mechanical result, the deviation of both visibilities goes to zero {\it always} 
when $\Omega t$ approaches $2\pi$ times an integer, which is the experimentally 
interesting region close to maximal visibility, cf. Fig.\,1. 

Since the visibility of Eq.\,(\ref{mod}) shows, for sufficiently 
short times ($\tau :=\Omega t\ll 1$, cf. Fig.\,\ref{mid}), a Gaussian decay, 
we may define the characteristic {\it decoherence time} $t_{CL}$ by: 
\begin{equation}\label{dectime} 
z_{CL}^2[1-\cos (\Omega t)]\approx  z_{CL}^2(\Omega t)^2/2 =:(t/t_{CL})^2/2 
\;\;,  \end{equation}  
and, correspondingly, for the case of a quantum mechanical mirror. This gives us 
the relevant decoherence times $t_{CL}=(z_{CL}\Omega )^{-1}$ and  
$t_{QM}=(z_{QM}\Omega )^{-1}$. Thus, we obtain the following relation:  
\begin{eqnarray}
t_{CL}&=&t_{QM}\cdot z_{QM}/z_{CL} 
\nonumber \\ [1ex] \label{decTrel} 
&=&t_{QM}
(\hbar\Omega /k_BT)^{1/2}\Big (\bar n(\hbar\Omega /k_BT)+1/2\Big )^{1/2}   
,\;\;\; \end{eqnarray} 
using Eq.\,(\ref{zrel}).   

In analogy to Eq.\,(\ref{hT}), we conclude here that the decoherence times coincide 
in the high-temperature limit. -- For experimentally relevant parameters \cite{Penrose},  
i.e., frequencies around   
$\Omega =2\pi\cdot 500$Hz, while maintaining $\kappa\approx 1$, and temperatures 
in the interval  $10^{-6}\mbox{K}\stackrel{<}{\sim}T\stackrel{<}{\sim}10^{-3}\mbox{K}$, 
such that 
$2.4\cdot 10^{-2}\stackrel{>}{\sim}\hbar\Omega /k_BT\stackrel{>}{\sim}2.4\cdot 10^{-5}$, 
we have that the  
discriminating factor $z_{QM}/z_{CL}$ in  
Eq.\,(\ref{decTrel}) deviates from 1 by less than $10^{-4}$. 
Therefore, the 
decoherence times $t_{CL}$ and $t_{QM}$ are the same to such accuracy   
that they will be difficult to distinguish experimentally, at present. 

For all practical purposes, the similitude of the features of mirror induced 
decoherence is a robust result, as we have 
demonstrated, considering both, either a classical mirror plus photon 
described by 
quantum-classical hybrid theory or a mirror plus photon described as fully 
quantum mechanical system \cite{Penrose}.   

\section{The probability to detect a photon}\label{PIP}
The mirror induced decoherence has been evaluated in terms of an off-diagonal matrix element 
of the reduced density operator for the photon. In this section, we relate this quantity 
to the experimentally accessible probability to find a photon, respectively, in one of 
the two detectors situated in the interferometer. 

They are given by:
\begin{equation}\label{P}
P_{i}\pton{t}=\mbox{Tr}\big (<\hat{\rho}\pton{t}>_{f}\hat{P}_{i})\;\;,\;\;\; i=1,2 
\;\;, \end{equation}
with the averaged density matrix given by:
\begin{eqnarray} \nonumber 
<\hat{\rho}\pton{t}>_{f}\;=\hskip 6cm
\\ [1ex] \nonumber 
\left(\begin{array}{cc} 
{\textstyle \frac{1}{2}}
&{\textstyle \frac{1}{2}}
\exp\big\{ -ik^2[\Omega t-\sin\pton{\Omega t}]
-z^2[1-\cos\pton{\Omega t}]\big \} 
\\ \mbox{c.c.}&
{\textstyle \frac{1}{2}}
\end{array}\right)
\\ ,\;\; \label{rhomatrix}   
\end{eqnarray}
and where $\hat P_{1}$ and $\hat P_{2}$ are projectors related to the two interferometer  
arms where the detectors are located \cite{Penrose}; ``c.c.'' denotes the complex 
conjugate of the upper off-diagonal matrix element. 
In the basis of $<\hat{\rho}\pton{t}>_{f}$ chosen here, the projectors are represented by:
\begin{equation}\label{P12}
\hat{P}_{1}=
\frac{1}{2}\left (\begin{array}{cc} 
1&1 \\ 1&1 \end{array}\right ) 
\;\;,\;\;\; 
\hat{P}_{2}=
\frac{1}{2}\left (\begin{array}{cc} 
1&-1 \\ -1&1 \end{array}\right )  
\;\;. \end{equation}
Inserting Eqs.\,$\pton{\ref{rhomatrix}}$ and $\pton{\ref{P12}}$ into Eq.\,$\pton{\ref{P}}$, 
we obtain:
\begin{eqnarray} 
P_{1,2}\pton{t}&=&\frac{1}{2}\big [1+2\mbox{Re}\pton{<\rho_{AB}\pton{t}>_{f}}\big ] 
\nonumber \\ [1ex] 
&=&\frac{1}{2}\big \{ 1\pm\cos\pqua{k^2\pton{\Omega t-\sin\pton{\Omega t}}} 
\nonumber \\ \label{prho} 
&\;&\cdot\exp\pqua{-z^2\pton{1-\cos\pton{\Omega t}}}\big \} 
\;\;. \end{eqnarray}
This presents an important relation, because it connects $<\hat{\rho}\pton{t}>_{f}$, the 
central quantity to learn about mirror induced decoherence, with the probability 
of observing a photon in one of the two detectors. In this way, in principle, we could 
learn about the former from experimental measurements of the latter. 

This result is independent of whether we consider the mirror as a classical or a quantum object, 
{\it i.e.} independent of whether we apply quantum theory to the whole interferometer set-up 
or the quantum-classical hybrid theory presently studied. 

\begin{figure}[!htpb]
\begin{center}
\includegraphics[width=7.3cm, height=7.5cm, keepaspectratio]{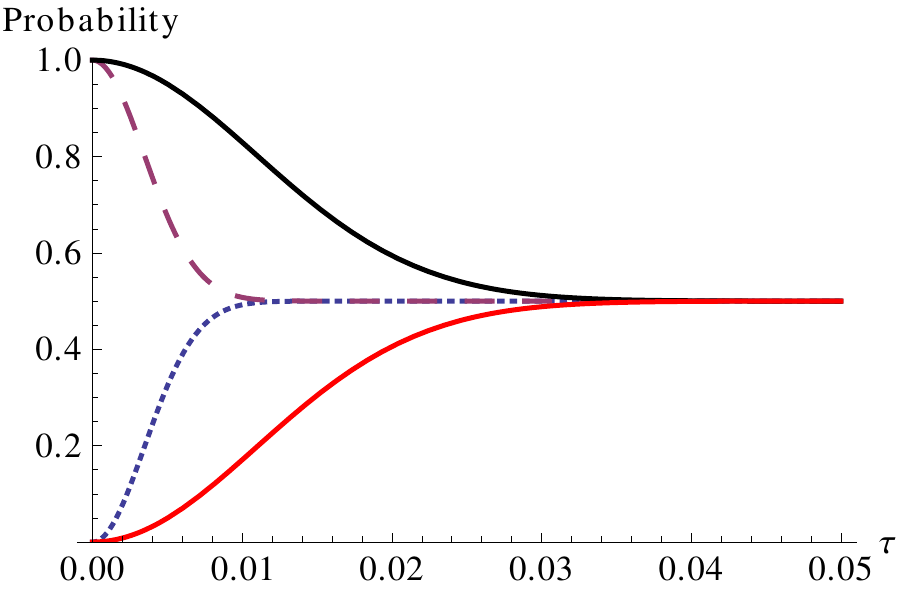}
\includegraphics[width=7.3cm, height=7.5cm, keepaspectratio]{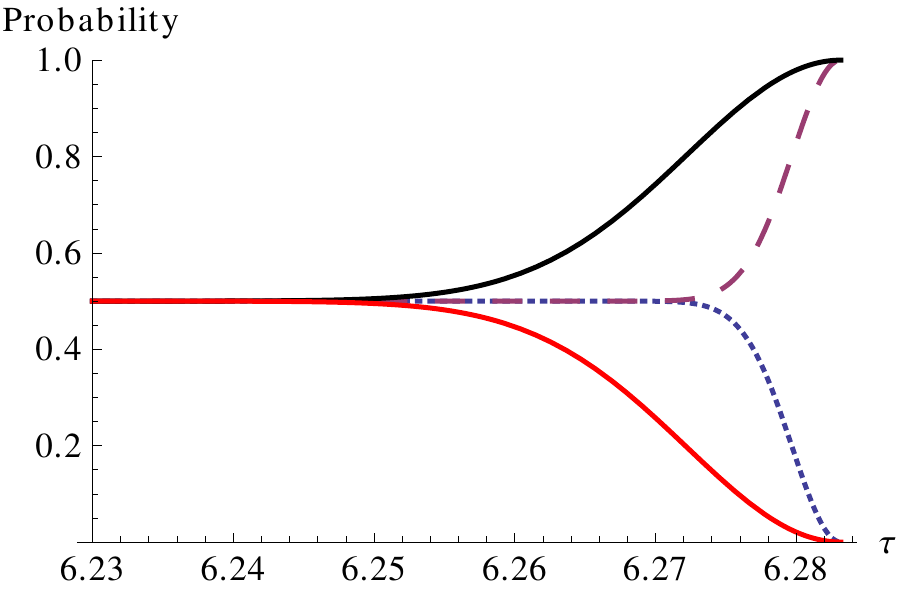}
\caption{The probabilities $P_{1,2}$ ($k=1$) as a function of $\tau :=\Omega t$, 
for several values of the temperature. 
The upper full line (black) and lower full line (red) represent, respectively, $P_{1}$ and $P_{2}$ for $T=10^{-4}$K;   
the dashed line (purple) and short-dashed line (blue) show, respectively, $P_{1}$ and $P_{2}$ 
for $T=10^{-3}$K.}
\end{center}
\end{figure}

\section{Conclusions and perspectives}
In this paper we have studied the optomechanical interferometer 
experiment of  {\it Marshall et al.} in the hybrid quantum-classical theory of  
Refs.\,\cite{Elze1, Elze2, Elze3,Elze4,Buric}.
Here, the mirror is considered as a perfectly classical rather than a quantum mechanical  
object and the quantum nature of the photon is preserved in a formally consistent 
framework.  

In Section\,$\ref{HH}$, we presented the hybrid Hamiltonian for the whole system, 
composed of the classical mirror and the quantum photon.
The Hamiltonian encodes all dynamical information regarding the system and has been 
employed to derive the corresponding equations of  motion. 
In Section\,$\ref{EM}$, we solved the equations analytically. 

In Section\,$\ref{MIDec}$, the solutions of the equations of motion have been used 
to obtain the off-diagonal elements of the reduced density matrix for the photon, 
which forms the starting point for our quantitative evaluation of the decoherence process 
induced by the classical mirror on the quantum photon. 
As in the fully quantum approach, this decoherence destroys interference effects 
and is detrimental to the formation of spatially separated coherent superposition states 
of the mesoscopic mirror.  

We emphasize that, according to the hybrid interaction scheme, the photon and the 
classical mirror presently do not become entangled. Thus, the mirror is at each 
moment of time in a classical pure state, unless thermal (or some other) fluctuations are explicitly introduced. Such classical fluctuations  play an analogous role here to the quantum 
fluctuations in the mirror state induced by entanglement if both parts are treated as 
quantum systems. 

More precisely, we have to distinguish two different cases. -- 
First, if the classical initial conditions of the mirror, namely its initial position and momentum, are exactly known, then no decoherence is observed: 
the photon remains in its initial pure state, the coherent superposition of being in either 
arm of the interferometer, and related interference effects are sustained. 
This clearly differs from the original quantum approach \cite{Penrose}, where also 
without thermal averaging over initial coherent oscillator states of the mirror, 
one finds mirror induced decoherence. 

This result leads us to conjecture that the absence of decoherence, when the initial 
conditions of the classical subsystem are completely fixed, is a general feature 
of a composite quantum-classical hybrid. A proof has to await future studies of these 
phenomena.  

Secondly, however, we have also examined the more realistic situation where some 
information about the classical initial conditions is lost and only a phase space 
probability distribution, instead, can be assumed or be experimentally prepared.

In particular, we have considered a thermal Boltzmann distribution specifying 
the mirror initial conditions. This leads to correspondingly averaged matrix elements 
of the reduced density matrix of the photon. Analyzing these, we find the surprising 
result that the mirror induced decoherence according to the hybrid theory essentially 
equals the one found in a fully quantum mechanical treatment \cite{Penrose}. 

This is nicely reflected in the corresponding decoherence timescales that we 
defined  and discussed  
in Section\,$\ref{MIDec}$, in particular for the experimentally relevant range of  temperatures.  
We pointed the stability of this equality 
with respect to variations of the physical parameters of the system (temperature $T$, 
mirror frequency $\Omega$, photon frequency $\omega$, cavity length $L$, and mirror mass $M$). 
We have found that the near-equality in the behaviour of the interferometer, whether 
treated as a quantum-classical hybrid or fully quantum mechanical system, is stable 
against such variations within the experimentally accessible regime. 

This extends to the experimentally measurable probability of finding a photon in one 
of the two detectors of the original interferometer arrangement \cite{Penrose}. 
In Section\,$\ref{PIP}$, we have related the off-diagonal matrix elements of the photon 
reduced density operator, from which mirror induced decoherence has always been calculated, 
to the probabilities to detect the photon in one of the two detectors.

An interesting study, which can also be performed on the basis of our formalism and results, 
will be to consider (thermally averaged) sqeezed initial states for a quantum mirror and 
correspondingly deformed (Boltzmann like) initial phase space distributions for a classical 
mirror. Will quantum theory and quantum-classical hybrid theory remain essentially 
indistinguishable also in this case, concerning mirror induced decoherence?   

In any case, our discussion may have implications for the interpretation of planned 
experiments, see, for example, Refs.\,\cite{BassiEtAl,YinEtAl,PaternostroEtAl} and 
further references therein. In fact, 
the optomechanical system of {\it Marshall et al.}, first of all, has 
been proposed to test various spontaneous wave function collapse models 
\cite{Diosi84,Diosi1,Penrose98,Bassi1,Diosi}. As indicated by our results, however, 
the system might not be suitable to discern a quantum from a classical mirror, 
given the accessible experimental parameters. In this case, the observation of ``anomalous  
decoherence'' ({\it i.e.}, when common sources of environmentally induced decoherence 
can be controlled) cannot unambiguosly be attributed to a rapid collapse mechanism, 
perhaps the mirror has been classical from the start and yet produces a similar 
decoherence signal.   

We conclude that applications of quantum-classical hybrid theory to describe presently 
considered experiments at the quantum-classical border, in particular when ``macroscopic'' 
components play a role, deserve further study. Last not least, since it is still  
thoroughly unknown whether, where, and what kind of border to expect.      

\section*{Acknowledgements}
It is a pleasure to thank L.~Di\'osi and F.~Giraldi    
for discussions on various occasions. H.-T.E. wishes to thank L.~Di\'osi also for kind 
hospitality and support during the Wigner-111 Symposium at Budapest, where 
part of this work was completed. 
A.L. and L.F. gratefully acknowledge support  
through Phd programs of their institutions; A.L. has been supported by      
an ERC AdG OSYRIS fellowship.



\begin{thebibliography}{99} 

\bibitem{Penrose}
W. Marshall, C. Simon, R. Penrose and D. Bouwmeester,
\emph{Towards quantum superpositions of a mirror},
Phys. Rev. Lett. {\bf 91}, 130401 (2003). 

\bibitem{Diosi84} L. Di\'osi, 
\emph{Gravitation and quantum-mechanical localization of macro-objects}, 
Phys. Lett. A {\bf 105}, 199 (1984).  

\bibitem{Diosi1} L. Di\'{o}si,
\emph{A universal master equation for the gravitational violation of quantum mechanics},
Phys. Lett. A {\bf 120}, 377 (1987). 

\bibitem{Penrose98} R. Penrose, 
\emph{Quantum computation, entanglement and state reduction} 
Phil. Trans. R. Soc. A {\bf 356}, 1927 (1998).  

\bibitem{Bassi1} S.L. Adler, A. Bassi and E. Ippoliti, 
\emph{Towards Quantum Superpositions of a Mirror: an Exact Open Systems 
Analysis - Calculational Details}, J. Phys. A {\bf 38}, 2715 (2005). 

\bibitem{Diosi} J. Bern\'{a}d, L. Di\'{o}si and T. Geszti, 
\emph{Quest for quantum superpositions of a mirror: high and moderately low temperatures}, Phys. Rev. Lett. {\bf 97,} 250404 (2006). 

\bibitem{decoherence} We use the term ``decoherence'' in the 
sense of diminishing off-diagonal elements of a density (sub)matrix, 
be it reversible (i.e., allowing ``recoherence'') or 
irreversible. The literature seems divided about reserving it only   
for the irreversible case, or not. 

\bibitem{Bose1} S. Bose, K. Jacobs and P.L. Knight, 
\emph{Preparation of non classical states in a cavity with a moving mirror},
Phys. Rev. A {\bf 56}, 4175 (1997). 

\bibitem{Bose2} S. Bose, K. Jacobs and P.L. Knight, 
\emph{Scheme to probe decoherence},
Phys. Rev. A {\bf 59}, 3204 (1999). 

\bibitem{Zurek} W.H. Zurek,
\emph{Decoherence, einselection and the quantum origins of the classical},
Rev. Mod. Phys. {\bf 75}, 715 (2003). 

\bibitem{MSR} M. Schlosshauer,
\emph{Decoherence, the measurement problems and interpretations of quantum mechanics},
Rev. Mod. Phys {\bf 76}, 1297 (2004). 

\bibitem{Sherry} T.N. Sherry and E.C.G. Sudarshan, 
\emph{Interaction between classical and quantum systems: A new approach to quantum measurement}, 
I. Phys. Rev. D {\bf 18}, 4580 (1978); {\it do.} II. {\it Phys. Rev.} D {\bf 20}, 857 (1979). 

\bibitem{Bou} W. Boucher and J. Trashen, 
\emph{Semiclassical physics and classical fluctuations}, 
Phys. Rev. D {\bf 37}, 3522 (1988). 

\bibitem{CaroSalcedo99} J. Caro and L.L. Salcedo, 
\emph{Impediments to mixing classical and quantum dynamics}, 
Phys. Rev. A {\bf 60}, 842 (1999).  

\bibitem{DiosiGisinStrunz} L. Di\'osi, N. Gisin and W.T. Strunz,   
\emph{Quantum approach to coupling classical and quantum dynamics}, 
Phys. Rev. A {\bf 61}, 022108 (2000).   

\bibitem{PeresTerno} A. Peres and D.R. Terno, 
\emph{Hybrid classical-quantum dynamics},   
Phys. Rev. A {\bf 63}, 022101 (2001).  

\bibitem{HallReginatto05} M.J.W. Hall and M. Reginatto, 
\emph{Interacting classical and quantum ensembles}, 
Phys. Rev. A {\bf 72}, 062109 (2005). 

\bibitem{ZhangWu06} Q. Zhang and B. Wu, 
\emph{General approach to quantum-classical hybrid systems and geometrical forces}, 
Phys. Rev. Lett. {\bf 97}, 190401 (2006). 

\bibitem{Hall08} M.J.W. Hall, 
\emph{Consistent classical and quantum mixed dynamics},  
Phys. Rev. A {\bf 78}, 042104 (2008).  

\bibitem{Manko12} V.N. Chernega, V.I. Man'ko, \emph{System with classical and 
quantum subsystems in tomographic probability representation},  
arXiv:1204.3854 (2012).      

\bibitem{Gnac} C.H. Chou, B.-L. Hu and Y. Suba\c{s}i, 
\emph{Macroscopic quantum phenomena from the large N perspective}, 
J. Phys.: Conf. Ser. {\bf 306}, 012002 (2011). 

\bibitem{Elze1} H.-T. Elze, 
\emph{Linear dynamics of quantum-classical hybrids},
Phys. Rev. A {\bf 85}, 052109 (2012). 

\bibitem{Elze2} H.-T. Elze, 
\emph{Four questions for quantum-classical hybrid theory},
J. Phys.: Conf. Ser. {\bf 361}, 012004 (2012). 

\bibitem{Elze3} H.-T. Elze, 
\emph{Proliferation of observables and measurement in quantum-classical hybrids}, 
Int. J. Qu. Inf. (IJQI) {\bf 10} No.\,8, 1241012 (2012). 

\bibitem{Elze4} H.-T. Elze, 
\emph{Quantum-classical hybrid dynamics: a summary},
J. Phys.: Conf. Ser. {\bf 442}, 012007 (2013). 

\bibitem{Buric} N. Buri\'c, I. Mendas, D.B. Popovi\'c, M. Radonji\'c and S. Prvanovi\'c, 
\emph{Statistical ensembles in the Hamiltonian formulation of hybrid quantum-classical 
systems}, 
Phys. Rev. A {\bf 86}, 034104 (2012).

\bibitem{Heslot} A. Heslot, 
\emph{Quantum mechanics as a classical theory},   
Phys. Rev. D {\bf 31}, 1341 (1985). 

\bibitem{Strocchi} F. Strocchi, 
\emph{Complex coordinates and quantum mechanics},   
Rev. Mod. Phys. {\bf 38}, 36 (1966). 

\bibitem{Man1} S. Mancini, V.I. Man'ko and P. Tombesi,
\emph{Ponderomotive control of quantum macroscopic coherence},
Phys. Rev. A {\bf 55}, 3042 (1997). 

\bibitem{Man2} S. Mancini, V.I. Man'ko and P. Tombesi,
\emph{Quantum noise reduction by radiation pressure},
Phys. Rev. A {\bf 49}, 4055 (1994). 

\bibitem{Pace} A.F. Pace, M.J. Collet and D.F. Walls,
\emph{Quantum noise reduction by radiation pressure},
Phys. Rev. A {\bf 47}, 3173 (1993). 

\bibitem{Law1} C.K. Law,
\emph{Interaction between a moving mirror and radiation pressure: a hamiltonian formulation},
Phys. Rev. A {\bf 51}, 2537 (1994). 

\bibitem{Law2} C.K. Law,
\emph{Effective hamiltonian for the radiation in a cavity with a moving mirror and 
a time varying dielectric medium},
Phys. Rev. A {\bf 49}, 433 (1993). 

\bibitem{Chen1} T. Hong, H. Yang, H. Miao and Y. Chen, 
\emph{Open quantum dynamics of single-photon optomechanical devices}, 
Phys. Rev. A {\bf 88}, 023812 (2013). 

\bibitem{Lorenzo}  L. Fratino, A. Lampo and H.-T. Elze, 
\emph{Entanglement dynamics in a quantum-classical hybrid of two q-bits and one 
oscillator}, Physica Scripta (2014), in press [arXiv:1408.1008]. 

\bibitem{Carlip} S. Carlip, 
\emph{Is quantum gravity necessary?}, 
Class. Qu. Grav. {\bf 25}, 154010 (2008).

\bibitem{BassiEtAl} A. Bassi, K. Lochan, S. Satin, T.P. Singh and H. Ulbricht, 
{\it Models of wave-function collapse, underlying theories, and experimental tests},
Rev. Mod. Phys. {\bf 85}, 2 (2013).

\bibitem{Chen2} H. Yang, H. Miao, D.-S. Lee, B. Helou and Y. Chen, 
\emph{Macroscopic quantum mechanics in a classical spacetime}, 
Phys. Rev. Lett. {\bf 110}, 170401 (2013). 

\bibitem{YinEtAl} Z. Yin, A.A. Geraci and T. Li,  
\emph{Optomechanics of Levitated Dielectric Particles}, 
Int. J. Mod. Phys. B {\bf 27}, 1330018 (2013).  

\bibitem{PaternostroEtAl} B. Rogers, N. Lo Gullo, G. De Chiara, G.M. Palma and 
M. Paternostro, 
\emph{Hybrid optomechanics for quantum technologies}, 
Quantum Measurements and Quantum Metrology {\bf 2}, 11 (2014).  

\end{thebibliography}
\end{document}